\documentclass{emulateapj}
\usepackage{apjfonts}

\newcommand{\Fig}[1]{Figure~\ref{#1}}
\newcommand{\Sec}[1]{\S~\ref{#1}}
\newcommand{\Tab}[1]{Table~\ref{#1}}

\newcommand{\imag}{\ensuremath{i^\prime}}
\newcommand{\zmag}{\ensuremath{z^\prime}}
\newcommand{\Msun}{\ensuremath{{\rm M}_\sun}}
\newcommand{\Zsun}{\ensuremath{{\rm Z}_\sun}}
\newcommand{\pow}[2]{\ensuremath{#1 \times 10^{#2}}}

\shorttitle{Ly$\alpha$ Emitting Galaxies at $z \sim 5.7$}
\shortauthors{Lai et al.}

\begin{document}

\newcommand{\lya}{Ly$\alpha$}
\newcommand{\zee}{z}

\title{The Stellar Population of \lya\ Emitting Galaxies at $\zee \sim 5.7$}
\author{Kamson Lai, Jia-Sheng Huang, Giovanni Fazio}
\affil{Harvard-Smithsonian Center for Astrophysics, 60 Garden Street,
  Cambridge, MA 02138, USA}
\email{klai@cfa.harvard.edu}
\and
\author{Lennox L.~Cowie, Esther M.~Hu, Yuko Kakazu}
\affil{Institute for Astronomy, University of Hawaii, 2680 Woodlawn
  Drive, Honolulu, HI 96822, USA}
\submitted{Accepted for publication in ApJ}

\begin{abstract}
We present a study of three Ly$\alpha$ emitting galaxies (LAEs),
selected via a narrow-band survey in the GOODS northern field, and
spectroscopically confirmed to have redshifts of $z \sim 5.65$.  Using
HST ACS and Spitzer IRAC data, we constrain the rest-frame
UV-to-optical spectral energy distributions (SEDs) of the galaxies.
Fitting stellar population synthesis models to the observed SEDs, we
find best-fit stellar populations with masses between $\sim 10^9 -
10^{10}$ \Msun\ and ages between $\sim 5 - 100$ Myr, assuming a simple
starburst star formation history.  However, stellar populations as old
as 700 Myr are admissible if a constant star formation rate model is
considered.  Very deep near-IR observations may help to narrow the
range of allowed models by providing extra constraints on the
rest-frame UV spectral slope.  Our narrow-band selected objects and
other IRAC-detected $z \sim 6$ \imag-dropout galaxies have similar 3.6
\micron\ magnitudes and $\zmag - [3.6]$ colors, suggesting that they
posses stellar populations of similar masses and ages.  This
similarity may be the result of a selection bias, since the
IRAC-detected LAEs and \imag-dropouts probably only sample the bright
end of the luminosity function.  On the other hand, our LAEs have blue
$\imag - \zmag$ colors compared to the \imag-dropouts, and would have
been missed by the \imag-dropout selection criterion.  A better
understanding of the overlap between the LAE and the \imag-dropout
populations is necessary in order to constrain the properties of the
overall high-redshift galaxy population, such as the total stellar
mass density at $z \sim 6$.
\end{abstract}

\keywords{cosmology: observations --- galaxies: evolution ---
  galaxies: high-redshift}

\section{Introduction} \label{Intro}

\begin{figure*}
\centering
\includegraphics[width=0.95\textwidth,bb=55 415 560 715]{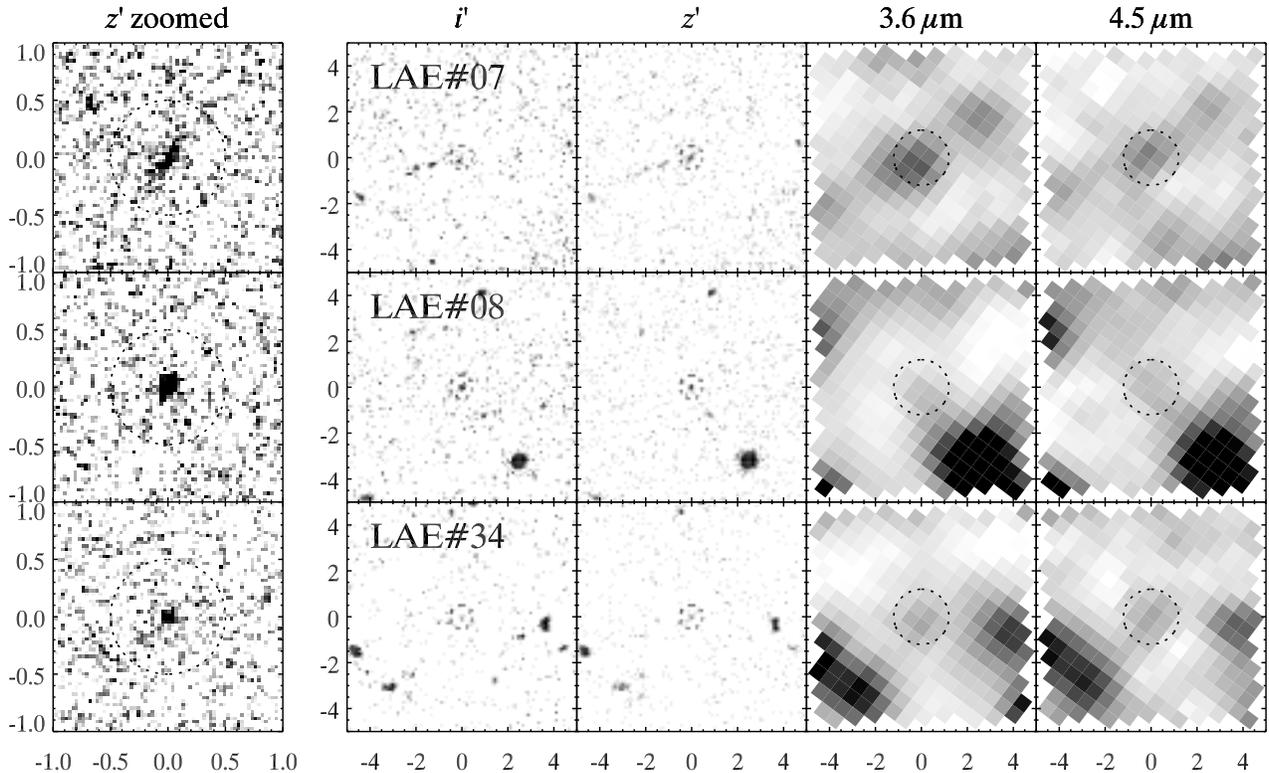}
\caption{Images of the three IRAC-detected $z \sim 5.65$ objects in
  \imag, \zmag, 3.6, and 4.5\micron.  Each panel is 10\arcsec across,
  except for the zoomed \zmag\ images, which are 2\arcsec across.  The
  images are rotated from the original mosaics, when necessary, so
  that north is up.  The ACS stamps are binned with a $3 \times 3$
  pixels box to bring out the faint extended sources.  The full
  resolution unbinned stamps are shown in the zoomed panels.  The
  circles indicate the apertures used in the photometry measurements.}
\label{stamps}
\end{figure*}

The star formation history of high-redshift galaxies is an important
problem in observational cosmology.  It provides tests for galaxy
formation theories \citep[e.g.][]{somer01, sprin03}, constrains the
sources of reionization \citep[e.g.][]{barka00, barka01}, and may even
shed light on the nature of the ``first stars''
\citep[e.g.][]{bromm04}.

Recently, the study of star formation at high redshift has been
catalyzed by the Infrared Array Camera (IRAC) on board the Spitzer
Space Telescope \citep{fazio04}.  The extraordinary sensitivity of
this instrument has enabled the detection of galaxies out to $z \sim
7$.  At such high redshifts, optical and near-IR observations
accessible to ground based telescopes and the HST sample the
rest-frame UV, which provides important information on the on-going
star formation and dust extinction.  On the other hand, IRAC samples
the rest-frame optical light, which is less susceptible to dust
extinction and is more representative of the emission from the less
massive stars making up the bulk of the stellar population.  Infrared
observations are therefore indispensable when studying the star
formation history of high-redshift galaxies.

Already, a sizable number of IRAC-detected $z \sim 6$ galaxies have
been discovered and studied in detail \citep{dow05, chary05, egami05,
eyles05, mobas05, schae05, yan05, mclur06, yan06}.  The infrared
observations provided by IRAC has allowed the galaxies' rest-frame
UV-to-optical spectral energy distributions (SEDs) to be constrained.
Estimates of the masses and ages of the galaxies' stellar populations
can then be obtained.  A stunning result from these studies is that
many of these high-redshift galaxies are massive ($\ga 10^{10}$
\Msun), and possess stellar populations that are older than several
hundred million years at a time when the age of the universe is less
than 1 Gyr.

The vast majority of the IRAC-detected $z \sim 6$ galaxies studied so
far have been found using the Lyman break technique \citep{steid95,
steid96, steid03} extended to high redshift \citep[the \imag-dropout
technique;][]{bouwe03, stanw03, yan04}.  Another, perhaps more
efficient, method to search for high-redshift galaxies is to select
candidates based on their Ly$\alpha$ emission \citep{rhoad01, ajiki03,
hu04, malho04, tanig05}.  The two methods suffer from different
selection biases.  For instance, while Ly$\alpha$ searches may allow
discovery of sources not detected in the continuum \citep{fynbo01},
they may preferentially select young galaxies in a dust-free
environment \citep{malho02}.  In a recent study, \citet{gawis06} found
that Ly$\alpha$ emitting galaxies at $z \sim 3$ may be younger and
less massive than Lyman break galaxies at similar redshifts.  It is
therefore interesting to compare the $z \sim 6$ Ly$\alpha$ selected
galaxies to the \imag-dropout sample.

In this paper, we present a study of three $z \sim 5.7$ galaxies
discovered in the Great Observatories Origins Deep Survey
\citep[GOODS;][]{dicki03} northern field via a narrow-band Ly$\alpha$
survey.  Using HST ACS and Spitzer IRAC data from GOODS, we study the
rest-frame UV and optical properties of Ly$\alpha$ emitting galaxies
at $z \sim 5.7$.  The goal is to derive stellar mass and age
estimates, and to identify differences, if any, compared to
\imag-dropout galaxies at similar redshifts.

The paper is organized as follows.  In \Sec{Data}, we describe the
candidate selection strategy and the photometry measurements.  In
\Sec{SED}, we present results from population synthesis modeling of
the observed SEDs.  We compare our sample of Ly$\alpha$ emitting
galaxies to the \imag-dropout sample in \Sec{CompLBG}, and the
conclusions are presented in \Sec{Con}.  Hereafter, we will refer to
the Ly$\alpha$ emitting galaxies as Ly$\alpha$ emitters (LAEs), and
the \imag-dropout galaxies simply as \imag-dropouts.  All magnitudes
quoted are in the AB magnitude system.  We adopt a cosmology of
$\{\Omega_m, \Omega_\Lambda, h\} = \{0.3, 0.7, 0.7\}$, consistent with
the recent results from WMAP \citep{sperg06}.

\section{Candidate Selection and Data Reduction} \label{Data}

\subsection{Candidate Selection and Follow-up Spectroscopy}

The high-redshift candidates are selected using a wide-field
narrow-band survey.  At the time of this paper's preparation, the
survey is being expanded to allow for deeper and more uniform
selection criteria over a wider region.  Details of the completed
survey, selection criteria, and catalog will be presented in an
upcoming paper (Hu et al., in prep.).  Here we will briefly summarize
the portion of the survey and the selection criteria used to construct
the current sample.

The narrow-band survey, carried out using SuprimeCam on the Subaru
Telescope, covers a 700 arcmin$^2$ area encompassing the GOODS
northern field.  This data set consists of broad-band continuum
$UVBRI\zmag$ images (\citealt{capak04}; Hu et al., in prep.),
supplemented by narrow-band observations taken with the NB816 filter
(Hu et al., in prep.).  The NB816 filter has a width of 120 \AA\ FWHM
and is centered around 8150 \AA, corresponding to Ly$\alpha$ at $z
\sim 5.7$.

The narrow-band observations allow for the selection of high-redshift
candidates based on the objects' narrow-band to broad-band flux
excess.  A detailed discussion of the general selection strategy for
$z \sim 5.7$ LAEs can be found in \citet{hu04}, while considerations
specific to this data set are discussed in Hu et al.\ (in prep.).  To
summarize, objects with $I-N > 0.7$ are selected down to a narrow-band
magnitude of $N = 25.5$.  Due to Ly$\alpha$ forest absorption and the
galaxies' intrinsic continuum break, the high-redshift candidates must
also satisfy either a) $R-\zmag > 1.8$ and be undetected in $B$ and
$V$, or b) be undetected in all passbands redward of $I$.

Candidates satisfying the selection criteria are followed up using the
DEIMOS spectrograph on the Keck Telescope.  At the DEIMOS resolution,
a number of candidates show single asymmetric broad lines, with long
red tails and steep blue drop-offs.  Such asymmetric lines are
characteristic of Ly$\alpha$ emission at high-redshift, and objects
displaying such emission lines are classified as $z \sim 5.7$ LAEs.

Our initial sample of spectroscopically confirmed $z \sim 5.7$ LAEs
consists of 20 objects (there are additional LAEs from the expanded
survey region), with 12 of them falling inside the $10\arcmin \times
16\farcm5$ region where deep IRAC observations are available from
GOODS.  Three objects, LAE\#07, LAE\#08, and LAE\#34, show significant
emission in the IRAC 3.6 and 4.5 \micron\ channels.  From now on, our
discussion will be focused on these three IRAC-detected $z \sim 5.7$
LAEs.  Spectroscopic redshifts of the objects are tabulated in
\Tab{FluxTab}.  Coordinates, spectra, and Ly$\alpha$ properties for
the objects will be presented in an upcoming paper (Hu et al., in
prep.) upon completion of the survey.

\subsection{X-Ray Nondetection}
In order to check for the possibility that our objects are
high-redshift AGNs, we turn to the 2 Ms Chandra Deep Field North
\citep[CDF-N;][]{alexa03}, in which all three of our objects lie.  We
cross-checked our objects with the point-source catalog by
\citet{alexa03} and found that none of our objects is detected in that
survey.  The three IRAC-detected objects lie near the edge of CDF-N,
where the sensitivity is lower at $\sim \pow{3}{-16}$ erg cm$^{-2}$
s$^{-1}$ (3$\sigma$, full-band 0.5 -- 8 keV; see Fig.~18 in
\citealt{alexa03}).  Nonetheless, a powerful AGN/quasar would have a
rest-frame 0.5 -- 8 keV luminosity of around $10^{44}$ erg s$^{-1}$
\citep{barge03, brand05}, which translates into an observed 0.5 -- 8
keV flux of \pow{4}{-16} erg cm$^{-2}$ s$^{-1}$ at $z = 5.65$
(assuming a photon index $\Gamma = 1.8$).  Therefore, if our sources
were x-ray luminous quasars we would expect to detect them.  However,
we cannot rule out the possibility of a weak AGN contribution.  On the
other hand, \citet{wang04} surveyed using Chandra a sample of 100 LAEs
at $z \sim 4.5$ found by the Large Area Lyman Alpha (LALA) survey, and
found no x-ray emission from any of the LAEs.  Stacking analysis also
reveals no detectable x-ray emission, placing a 3$\sigma$ upper limit
on the average 2 -- 8 keV luminosity at $L < \pow{2.8}{42}$ erg
s$^{-1}$.  Our objects are selected in a similar manner as the LALA
LAEs, and so should be higher redshift counterparts of the LALA
objects.  Therefore, given the objects' nondetection in CDF-N and the
\citet{wang04} results, we consider it unlikely that our objects are
high-redshift AGNs.

\subsection{Photometry}

\begin{figure}
\plotone{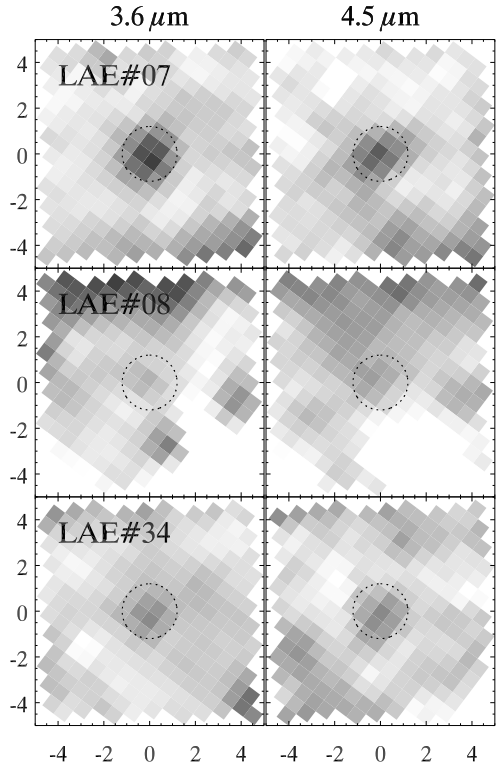}
\caption{Object images in the IRAC 3.6 and 4.5 \micron\ bands with the
  neighboring objects subtracted.}
\label{sub_stamps}
\end{figure}

\begin{deluxetable*}{lcr@{$\pm$}lr@{$\pm$}lr@{$\pm$}lr@{$\pm$}l}
\tablewidth{0pt}
\tablecaption{Measured Flux Densities \label{FluxTab}}
\tablehead{ &
  \colhead{Redshift} &
  \multicolumn{2}{c}{\imag}      & \multicolumn{2}{c}{\zmag}       & 
  \multicolumn{2}{c}{3.6 \micron} & \multicolumn{2}{c}{4.5 \micron}}
\startdata
LAE\#07 & 5.635 &  0.071 &  0.047 &   0.135 
  &  0.030 &   0.53 &   0.20 &   0.41 &   0.17 \\ 
LAE\#08 & 5.640 &  0.070 &  0.045 &   0.124 
  &  0.033 &   0.51 &   0.20 &   0.61 &   0.17 \\
LAE\#34 & 5.671 &  0.073 &  0.043 &  0.091 
  &  0.025 &   0.31 &   0.20 &   0.28 &   0.17
\enddata
\tablecomments{Flux densities are given in units of $\mu$Jy.  Quoted
  errors are the 1$\sigma$ uncertainties.  In the IRAC bands, the
  errors include systematic uncertainties coming from confusion with
  neighboring sources.  The objects are undetected in $B$, $V$, $R$,
  5.8 \micron, and 8.0 \micron, with 3$\sigma$ upper limits of 0.042,
  0.036, 0.063, 1.0, and 1.0 $\mu$Jy, respectively.}
\end{deluxetable*}

The optical photometry is measured from the v1.0 release of the HST
ACS data from GOODS.\footnotemark\ A detailed description of the GOODS
ACS data is given in \citet{giava04}.  To summarize, the northern
field of GOODS covers an area of $\sim 160$ arcmin$^2$ in four
passbands: F435W ($B$), F606W ($V$), F775W (\imag) and F850LP
(\zmag).  The images have PSF FWHM $\sim 0\farcs12$ and reach depths
of 27.4, 27.5, 26.8, and 26.5 mag ($3\sigma$ in 1\arcsec-diameter
apertures) in the $B$, $V$, \imag, and \zmag bands, respectively.
\footnotetext{\raggedright http://archive.stsci.edu/pub/hlsp/goods/v1/
h\_goods\_v1.0\_rdm.html}

Infrared coverage is provided by deep Spitzer IRAC observations,
covering the same field as the ACS imaging, obtained as part of the
GOODS program (Dickinson et al., in prep.).\footnotemark\ The data
consist of imaging in four passbands, centered at 3.6, 4.5, 5.8 and
8.0 \micron.  The IRAC photometry are measured using images from our
own independent reduction of the GOODS data.  The $3\sigma$ magnitude
limits (in 2\farcs4-diameter apertures) in our images are 26.2, 26.4,
25.0, and 25.1 mag in the 3.6, 4.5, 5.8, and 8.0 \micron\ bands,
respectively.  The PSF FWHM is $\sim 2\arcsec$ on average, but is
slightly better at $\sim 1\farcs8$ in the 3.6 and 4.5 \micron\ bands.
\footnotetext{\raggedright
http://data.spitzer.caltech.edu/popular/goods/Documents/
goods\_dataproducts.html}

The three objects are detected only in the \imag, \zmag, 3.6 \micron,
and 4.5 \micron\ bands.  Stamp images of the objects in these four
passbands are shown in \Fig{stamps}.  LAE\#07 and LAE\#08 appear to be
marginally resolved in the ACS images, with approximate angular sizes
of $\sim 0\farcs7$ and 0\farcs35, respectively.  Taken at face value,
the angular sizes correspond to physical sizes of $\sim$ 4 and 2 kpc
for LAE\#07 and LAE\#08.

We perform aperture photometry on the ACS images for each of the three
objects.  Because of the extended nature of the objects in the ACS
images, extra care has to be taken when selecting the appropriate
aperture size.  We adopt a 1\arcsec-diameter aperture in the
measurements.  Curve of growth analysis of the three objects suggests
that $\la 1\%$ of the total flux fall outside an 1\arcsec-diameter
aperture.  While a large aperture is not ideal for minimizing
background noise contributions, it is necessary in order to measure
the total flux of the objects.  Since the fraction of enclosed flux
within our chosen aperture is close to unity, and the objects are
intrinsically faint and extended, we opted not to apply aperture
corrections to the photometry.

Photometry measurements on the IRAC images are considerably more
challenging.  At the IRAC resolution, all three objects are unresolved
and can be treated as point sources.  However, because of the large
PSF, neighboring objects may significantly contaminate the photometry.
For example, LAE\#07 is heavily blended with two other faint
neighbors, while LAE\#08 is located in the wings of a bright
foreground galaxy (Fig.~\ref{stamps}).

To address this problem, we use a deblending technique whereby the
contaminating neighbors are subtracted using the high resolution ACS
images convolved with the IRAC PSF.  We use the ACS \zmag-band image
as input since it is the closest to the IRAC bands in wavelength and
the objects are brightest in this passband.  In fact, we in general
use the \zmag-band image as the reference for object positions.  The
IRAC PSF is obtained directly from the images by stacking $\sim 10$
bright point sources within the field.

For each contaminating object in the vicinity of the target, the
\zmag-band image is convolved with the IRAC PSF.  Then, after first
subtracting all the neighboring sources using the current best-fits,
the convolved image of the contaminating object is fitted to its IRAC
counterpart.  The parameters fitted for are the object's amplitude,
background, and position.  This process is iterated until a converged
solution for every contaminating object is found.  Using the best-fit
solution, the contaminating objects are subtracted from the original
IRAC image, leaving only the target.  The result of this deblending
procedure is shown in \Fig{sub_stamps}.  It is important to note that
this deblending procedure works best when the objects to be subtracted
are unresolved, such as the neighbors of LAE\#07.  The procedure is
less successful in subtracting the bright extended foreground galaxy
neighboring LAE\#08, most likely due to uncertainties and spatial
variations in the IRAC PSF, and possible wavelength dependence in the
foreground galaxy's morphology.  Nonetheless, \Fig{sub_stamps} shows
that the deblending procedure works quite well in removing
contamination from neighboring sources.

After the contaminating neighbors are subtracted from the IRAC images,
we measure the objects' photometry in 2\farcs4-diameter apertures,
centered at the positions derived from the \zmag-band images.  The
aperture size is chosen to be as small as possible to minimize
background noise and residual flux from nearby objects.  Aperture
corrections factors of $\sim 0.5$ are applied to the measured fluxes.

Uncertainties in the ACS photometry are estimated using a Monte-Carlo
procedure.  Artificial point sources, with fluxes equal to the
objects, are inserted into the images.  The positions of the
artificial sources are assigned randomly apart from the requirement
that they are at least 1 aperture diameter away from detected sources.
We then apply the same measurement procedure we used on the objects to
the artificial sources.  The resulting dispersion in the measured
fluxes will serve as an estimate of the uncertainties in the objects'
photometry.  This procedure provides an estimate of the errors
introduced by sky fluctuations and sky subtraction, and it takes into
account the correlations between adjacent pixels introduced by the
drizzle procedure used to produce the ACS mosaics.  The errors derived
in this way also include the confusion noise due to faint undetected
sources.

For the IRAC images, there is an additional source of error coming
from the neighbor subtraction.  Ideally, we would apply a similar
Monte-Carlo procedure, repeating the neighbor subtraction process many
times on artificial sources to obtain the error estimates.  However,
the convolution-based neighbor subtraction process is time-consuming,
and so a full Monte-Carlo approach is impractical.

We therefore try to mimic the effects of neighbor subtraction using
the PSF-fitting package StarFinder \citep{diola00}.  The errors in the
source subtraction are dominated by image quality issues, such as
variations of the PSF across the field, and by the properties of the
individual sources (e.g.\ deviations from being perfect point
sources).  The differences in the source subtraction using StarFinder
and our convolution-based procedure should be small compared to the
aforementioned uncertainties.  Hence, we believe that performing the
source subtraction using StarFinder will give a reasonable estimate of
the errors introduced by our convolution-based neighbor subtraction
process.

Using StarFinder, sources are detected and subtracted from the IRAC
images.  Artificial sources are inserted randomly into the source
subtracted image and their fluxes are then measured.  Any errors
arising from the imperfect source subtractions, as well as the
statistical errors coming from the background and its subtraction,
would be included in the resulting measured flux dispersion.  The IRAC
photometry errors thus derived are conservative since we expect our
convolution-based subtraction to perform better than StarFinder in the
case where the objects are not perfect point sources.  Also, the
artificial sources are placed purely randomly throughout the image, so
they may overlap subtracted sources.  This tends to overestimate the
uncertainties since the errors in the subtraction are larger near the
center of the sources where the flux is higher.  The results of the
flux measurements and error estimates described above are summarized
in \Tab{FluxTab}.

Finally, we supplement the ACS and IRAC imaging using data from the
Hawaii-HDF-N project \citep{capak04}.  We attempt to measure the
objects' photometry in the $R$-band images from Subaru and the {\it
HK}$^\prime$ images from the UH 2.2m telescope.  However, these images
are shallower than the ACS and IRAC data, and none of the objects are
detected (in 3\arcsec-diameter apertures) in either $R$ or {\it
HK}$^\prime$.  We include the $R$-band upper limit (27.2 mag,
3$\sigma$) in our analysis for completeness.  The {\it
HK}$^\prime$-band upper limit (22.7 mag, 3$\sigma$) is too high to
make a difference in the following analysis, and is therefore omitted.

\section{Spectral Energy Distribution and Stellar Population} \label{SED}

\begin{figure}
\plotone{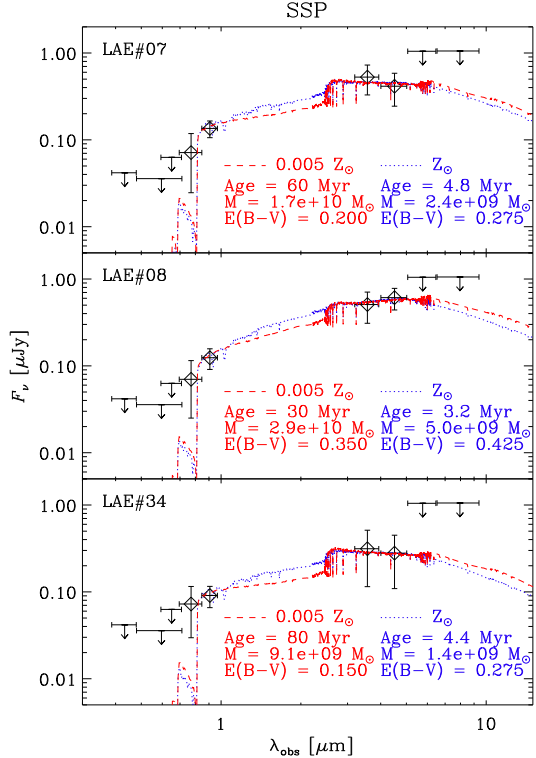}
\caption{Best fit SSP models to the three IRAC-detected $z \sim 5.7$
  LAEs.  The blue-dotted (red-dashed) line is for a model with $Z =
  \Zsun$ (0.005 \Zsun).  The observed SEDs are plotted as diamonds
  with error bars.  The horizontal error bars indicate approximately
  the FWHM of the passbands.  When an object is undetected in a
  certain passband, the corresponding 3$\sigma$ upper limit is shown.
  Note that a 30\% fractional error is added to the \imag-band to
  account for the Ly$\alpha$ line contribution.}
\label{SspSED}
\end{figure}

\begin{figure}
\plotone{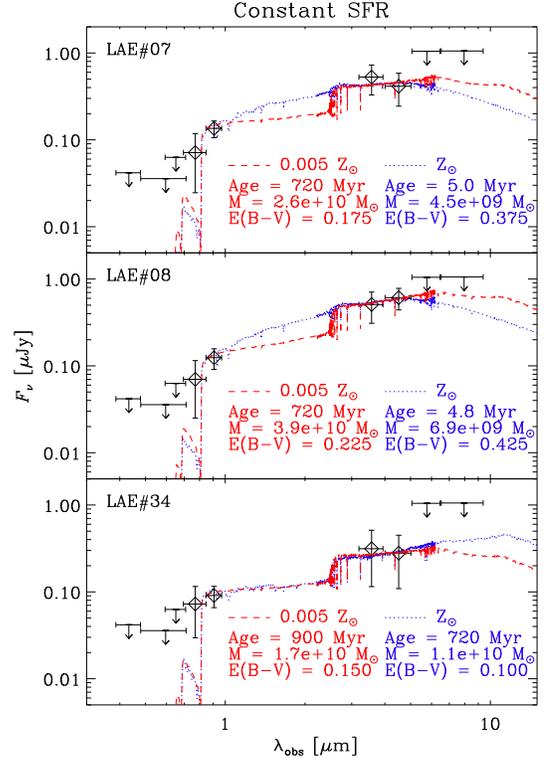}
\caption{Best fit constant SFR models to the three IRAC-detected $z
  \sim 5.7$ LAEs.  Symbols and line styles have the same meaning as in
  \Fig{SspSED}.}
\label{ConSED}
\end{figure}

\subsection{Population Synthesis Modeling} \label{Model}
After we have obtained reliable photometry for the galaxies, we use
the stellar population synthesis model of \citet[][hereafter
BC03]{bc03} to model the galaxies' SEDs.  The basic strategy is to
calculate model SEDs, varying the mass, age, dust reddening,
metallicity, and star formation history.  The model SEDs are
redshifted and integrated through the filter response function.  The
model predictions are then compared to the observed photometry and the
best-fitting model is found by $\chi^2$ minimization.

For simplicity, and to facilitate comparison with other works in the
literature, we will assume throughout this study the \citet{salpe55}
initial mass function (IMF), implemented in the BC03 models with lower
and upper mass cutoffs of 0.1 and 100 \Msun.  We will discuss the
effects of using a different IMF and other variations in the
population synthesis models in \Sec{AltMod}.  We explore models with
five different metallicities: Z/\Zsun\ = 0.005, 0.02, 0.2, 0.4, and 1.
Note that metallicity is not treated as a free parameter in the
fitting process, but rather an input assumption of the model.  The
effect of dust reddening is included using the \citet{calze00} model,
with $E(B-V)$ varying between 0.0 -- 1.0.  The effect of Ly$\alpha$
forest attenuation is included using the model of \citet{madau95}.
There is no need to fit for the objects' redshifts, since these are
known from spectroscopy.  An additional constraint arising from the
known redshifts is that the inferred ages should not exceed the age of
the universe, which is approximately 0.99 Gyr at $z \sim 5.65$.  We
therefore impose an upper limit of 0.9 Gyr on the age of the best-fit
models.

The star formation history (SFH) is parameterized by the time
evolution of the star formation rate (SFR).  We consider two SFHs in
our analysis: Simple Stellar Population (SSP), i.e.\ instantaneous
burst, and constant SFR.  We choose not to consider more complicated
SFHs, such as an exponentially declining SFR, because our ability to
constrain extra parameters is limited.  All our objects are detected
in only four passbands, so we can constrain up to three parameters,
which we choose to be mass, age, and dust reddening $E(B-V)$.  Using
an exponential SFR would introduce an extra parameter, the star
formation timescale $\tau$, which can be constrained only if we fix
the value of another parameter such as $E(B-V)$.  We will return to
this point later in \Sec{AltMod}.  Note that the SFHs used in our
analysis (SSP and constant SFR) represent the two extremes in star
formation timescales, and an exponential SFR can be thought of as an
intermediate between these two examples.

With two SFHs and five metallicities, there are 10 different classes
of models in total.  The best-fitting model within each class is found
by $\chi^2$ minimization.  There are three free parameters (mass, age,
and reddening) and four data points (\imag, \zmag, 3.6 \micron, and
4.5 \micron), resulting in 1 degree of freedom in the $\chi^2$ fits.
In practice, a commonly used measure of the goodness of fit is
$\chi_\nu^2$, the $\chi^2$ per degree of freedom.  When $\nu = 1$, a
model is rejected at the 1, 2, and 3-$\sigma$ levels if $\chi_\nu^2$
is larger than 1, 4, and 9, respectively.  We in general consider a
fit to be acceptable if it gives $\chi_\nu^2 < 4$.  In certain cases
when $\nu = 2$ (see \Sec{Fit}), a fit is deemed acceptable if
$\chi_\nu^2 < 3.09$ (the corresponding 2$\sigma$ level for $\nu = 2$).

In order to make full use of the data available, we incorporate
undetected data points as follows.  Given the detection limit of the
exposure $\sigma$ and the model flux in the channel $\mu$, we
calculate the probability $I = P(<\sigma|\mu)$ that the actual flux of
the object is less than $\sigma$.  For the purpose of this
calculation, the flux is assumed to be normally distributed with mean
$\mu$ and width $\sigma$.  For each undetected channel, a value of
$-\ln(I)$ is then added to the total $\chi^2$.  In this way, if the
model flux is much higher than the detection limit, $\mu \gg \sigma$,
$I$ would be small and the model would incur a large $\chi^2$ penalty.
For example, if the model flux in a certain channel is at the
$3\sigma$ detection limit, $\mu = 3\sigma$, the $\chi^2$ penalty would
be 3.8.  In practice, the undetected data points have only a small
effect on the total $\chi^2$, since the detection limits in most
channels can easily be satisfied by reasonable models.

Since the Ly$\alpha$ emission line is located inside the \imag-band at
$z \sim 6$, and the BC03 model does not predict emission line
strengths, we have to take into account the possible Ly$\alpha$ line
emission when fitting to the data.  The fractional contribution of the
line to the broad-band flux can be approximated as $F_L/F_B =
(R-1)W_N/W_B$, where $W_N/W_B$ and $R = F_{\nu}(N) / F_{\nu}(B)$ are
the ratios of the narrow-band to broad-band widths and flux densities,
respectively.  Our objects are selected based on their narrow-band to
broad-band flux excess, and typically have $N-I \sim -1.5$.
Substituting $W_N \approx 120$ \AA\ and $W_B \approx 1400$ \AA\ for
the NB816 and $I$ filters, and $R = 10^{-0.4(N-I)} \approx 4$, we find
that the Ly$\alpha$ line may contribute around 30\% of the total
$I$-band flux.  Note that this approximation implicitly assumes that
the Ly$\alpha$ forest truncates similar fractions of the narrow-band
and broad-band fluxes.  Since Ly$\alpha$ at $z \sim 5.7$ is located
slightly redward of the \imag-band center, Ly$\alpha$ forest
truncation affects the \imag-band flux more severely than $I$ or
NB816.  The Ly$\alpha$ line contribution may therefore be substantially
higher than 30\% in the \imag-band.  Nevertheless, we add an
optimistic 30\% fractional error to the \imag-band flux when we
perform the $\chi^2$ fits.

\begin{figure}
\plotone{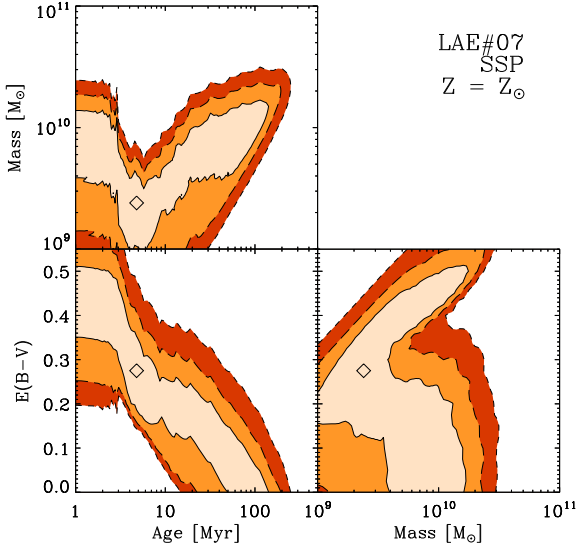}
\caption{$\Delta\chi^2$ contour for LAE\#07 in the \Zsun\ SSP model.
  The diamond marks the best fitting model.  The solid, long-dashed,
  and short-dashed lines represent the 1, 2, and 3-$\sigma$ confidence
  regions.}
\label{contour}
\end{figure}

\subsection{Fitting Results} \label{Fit}

\begin{figure}
\plotone{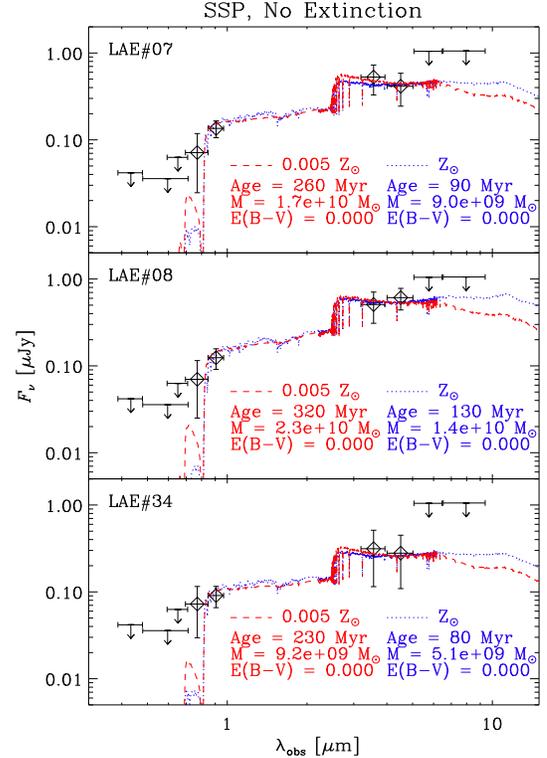}
\caption{Best fit SSP models to the three IRAC-detected $z \sim 5.7$
  LAEs with fixed $E(B-V) = 0$.  Symbols and line styles have the same
  meaning as in \Fig{SspSED}.}
\label{SspSEDnodust}
\end{figure}

\begin{deluxetable*}{lcr@{$\times$}lccccr@{$\times$}lcc}
\tabletypesize{\scriptsize}
\tablewidth{0pt}
\tablecaption{Best-Fit Model Parameters \label{ParTab}}
\tablehead{ &
  \colhead{Age [Myr]} & \multicolumn{2}{c}{Mass [\Msun]} & 
  \colhead{$E(B-V)$}  & \colhead{$\chi_\nu^2$} &
  \hspace{24pt} &
  \colhead{Age [Myr]} & \multicolumn{2}{c}{Mass [\Msun]} &
  \colhead{$E(B-V)$}  & \colhead{$\chi_\nu^2$}}
\startdata
& \multicolumn{5}{c}{SSP, \Zsun} & & \multicolumn{5}{c}{SSP, 0.005 \Zsun} \\
\hline
LAE\#07 & 4.8 & 2.4 & $10^{ 9}$ & 0.275 & 2.05 & &  60 & 1.7 & $10^{10}$ & 0.200 & 1.97 \\
LAE\#08 & 3.2 & 5.0 & $10^{ 9}$ & 0.425 & 2.19 & &  30 & 2.9 & $10^{10}$ & 0.350 & 2.20 \\
LAE\#34 & 4.4 & 1.4 & $10^{ 9}$ & 0.275 & 2.42 & &  80 & 9.1 & $10^{ 9}$ & 0.150 & 2.36 \\
\hline
& \multicolumn{5}{c}{Constant SFR, \Zsun} & & 
\multicolumn{5}{c}{Constant SFR, 0.005 \Zsun} \\
\hline
LAE\#07 & 5.0 & 4.5 & $10^{ 9}$ & 0.375 & 2.14 & & 720 & 2.6 & $10^{10}$ & 0.175 & 2.19 \\
LAE\#08 & 4.8 & 6.9 & $10^{ 9}$ & 0.425 & 2.19 & & 720 & 3.9 & $10^{10}$ & 0.225 & 2.18 \\
LAE\#34 & 720 & 1.1 & $10^{10}$ & 0.100 & 2.47 & & 900 & 1.7 & $10^{10}$ & 0.150 & 2.37 \\
\hline
& \multicolumn{5}{c}{SSP, \Zsun, No Extinction} & & 
\multicolumn{5}{c}{SSP, 0.005 \Zsun, No Extinction} \\
\hline
LAE\#07 &  90 & 9.0 & $10^{ 9}$ & 0.000 & 1.25 & & 260 & 1.7 & $10^{10}$ & 0.000 & 1.05 \\
LAE\#08 & 130 & 1.4 & $10^{10}$ & 0.000 & 1.45 & & 320 & 2.3 & $10^{10}$ & 0.000 & 1.28 \\
LAE\#34 &  80 & 5.1 & $10^{ 9}$ & 0.000 & 1.41 & & 230 & 9.2 & $10^{ 9}$ & 0.000 & 1.27
\enddata
\end{deluxetable*}

The observed SEDs of the three galaxies are shown in \Fig{SspSED}.
Ignoring the lines for the moment, the first thing to take note is
that all three galaxies exhibit similar rest-frame UV and optical
emission properties.  This is not surprising since the three objects
are selected from same data set using the same selection criteria.
The galaxies' SEDs are characterized by red UV to optical colors, with
the flux increasing by more than a factor of 3 from \zmag\ to 3.6
\micron.  The SEDs also show signs of flattening out beyond 3.6
\micron.

The best fit BC03 models to the observed SEDs are shown in
\Fig{SspSED} and \Fig{ConSED} for the SSP and constant SFR models,
respectively.  Only models with the two extreme metallicities (\Zsun\
and 0.005 \Zsun) are shown, even though fits were performed for all
five possible metallicities.  In practice, we find good fits with
competitive $\chi^2$ for all five metallicities.  Note that the
\imag-band flux is consistently higher than the models, owing to the
Ly$\alpha$ line emission.  Also, since the \imag-band straddles the
Lyman-break, the model band-integrated flux is larger than it seems,
and is usually within 1$\sigma$ of the observed value.

The best-fit parameters differ depending on the SFH and metallicity
(\Tab{ParTab}).  The best-fit \Zsun\ SSP models to the three LAEs have
ages $\sim 5$ Myr, masses $\sim \pow{5}{9}$ \Msun, and $E(B-V) \sim
0.3 - 0.4$.  The best-fit 0.005 \Zsun\ SSP models favor slightly older
ages ($\sim 50$ Myr), larger masses ($\sim 10^{10}$ \Msun), and
somewhat lower extinction.  This is mostly because metal-poor stars
produce more UV photons, so the best-fit models need to be older in
order to match the red UV-to-optical colors in the data.  These
results suggest that our sample of $z \sim 5.7$ LAEs displays
qualities of dusty young galaxies, seen immediately after or during a
burst of star formation.

Compared to the SSP models, the constant SFR models in general yield
older ages at $\ga 700$ Myr and slightly larger masses at a few times
$10^{10}$ \Msun.  In some cases, however, the best-fit constant SFR
models have similar parameter values as the SSP models.  This is not
surprising since for young ages the constant SFR model is similar to
the SSP model.

Given that we only have four data points, some of which have fairly
large error bars, it is important to explore the range of models
allowed by the data.  Furthermore, the somewhat high levels of
extinction we obtained may seem be a little surprising.  Several
previous studies have found little or no dust extinction in $z \sim 6$
\imag-dropouts \citep{dow05, eyles05, mobas05, yan05}.  The fact that
our galaxies are selected based on their Ly$\alpha$ emission would
also lead one to expect low extinction values, since Ly$\alpha$
photons are very susceptible to dust scattering \citep{charl91,
chen94, hanse06}.  Do the data permit models with low extinction?
What is the range of age and mass allowed by the data?

To answer these questions, we calculate confidence regions for the
best-fit parameters.  In \Fig{contour}, we show the $\Delta\chi^2$
contours for LAE\#07 in the \Zsun\ SSP model.  It can be seen from the
figure that firm upper limits on the age and mass can be placed on
LAE\#07.  In the context of the SSP model, LAE\#07 has a mass of $\la
\pow{3}{10}$ \Msun, and an age of $\la 300$ Myr.  The contours and
limits for the other two objects are similar.

One important thing to note in \Fig{contour} is that there is a range
of models that fits the data.  This is the result of a fundamental
degeneracy in the models: the red UV-to-optical colors of the observed
SEDs can be satisfied by either a significant Balmer break (implying a
relatively mature stellar population), or a young stellar population
with high extinction.  \Fig{contour} shows that even though the
best-fit model has a young age and high extinction, there are models
with lower extinction values and older ages that will fit the data
equally well.

Because of this degeneracy, we performed an alternative fit to the
data forcing $E(B-V) = 0$.  The results of this fit for the SSP model
are presented in \Fig{SspSEDnodust}.  Good fits to the data can be
obtained for all metallicities.  The best-fit models to the three
galaxies in the \Zsun\ case have masses $\sim 10^{10}$ \Msun, and ages
$\sim 100$ Myr.  The 0.005 \Zsun\ model gives older ages (around a few
hundred Myr) and slightly larger masses.  In general, the no
extinction models have a lower $\chi_\nu^2$ (see \Tab{ParTab}), mostly
due to the extra degree of freedom ($\nu = 2$ in this case).  Note
that in the SSP models presented here, there will not be enough young
massive stars around to produce significant Ly$\alpha$ emission
becuase of the older ages.  The results we obtained should therefore
be regarded as a fit to the older stellar population that dominates
the total mass of the system.  The observed Ly$\alpha$ emission can
still be explained by the presence of a young but significantly less
massive stellar population within the galaxy.  We find no satisfactory
constant SFR models with $E(B-V) = 0$.  Without dust extinction, these
models produce too much UV compared to the data.

In a recent study, \citet{ledel06} made detailed predictions of the
properties of LAEs, based on a hierarchical galaxy formation model.
We find that the best-fit stellar masses of our LAEs are about 1 -- 2
orders of magnitude larger that the values predicted by
\citet{ledel06}.  The most probable reason for this discrepancy is
that a top heavy IMF is used in \citet{ledel06}, resulting in a lower
M/L ratio and hence lower overall mass.  By requiring that our LAEs be
detected by IRAC, we may also be selecting galaxies from the high-mass
end of the distribution (\Sec{comp_sp}).  Other properties of the
LAEs, such as the broad-band magnitudes as a function Ly$\alpha$ flux,
agree quite well with the predictions of the \citet{ledel06} model.

In summary, we find that the three IRAC-detected $z \sim 5.7$ LAEs in
our sample show similar emission properties and can be fitted with
similar model SEDs.  The current data are unable to distinguish
between models with different metallicities, as best-fit models with
competitive $\chi^2$ can be found for all metallicities.  The observed
SEDs are broadly consistent with two classes of models.  In one model,
the galaxies are dusty, young (ages $\la 100$ Myr), and relatively
less massive ($M \sim 10^9$ \Msun).  In the other model, the galaxies
are less dusty, more massive ($M \sim 10^{10}$ \Msun), and older ($\ga
100$ Myr) with a significant Balmer break.  Both types of models can
provide satisfactory fits to the data.  This degeneracy stems from the
fact that we only have two data points (\imag\ and \zmag) with which
to measure the UV spectral slope, which is sensitive to both dust
extinction and metallicity.  Very deep near-IR observations in the
$J$, $H$, or $K$ bands would provide valuable constraints on the UV
spectral slope, and may help to break the degeneracy and rule out some
of the possible models.

\subsection{Alternative Models} \label{AltMod}
In addition to the basic models discussed in the previous section, we
also tried to fit a number of alternative models to the data, in order
to explore several effects not included in the basic fits.  These
alternative models are discussed in turn below.

{\em Exponential SFR.} --- We did not consider an exponential SFR in
our basic fits because with only four data points, it is not possible
to fit for the mass, age, reddening, and star formation timescale
($\tau$) simultaneously.  However, if we are willing to sacrifice one
degree of freedom in another parameter, we can attempt to fit an
exponential SFR model to the data.  In the previous section, we found
that the galaxies can be fitted by models with very little dust
extinction.  Other studies of galaxies at similar redshifts also found
little or no dust extinction \citep{dow05, eyles05, mobas05, yan05}.
We therefore attempt to fit an exponential SFR model with $E(B-V)$
fixed at zero.  Good fits to the data can be found using all
metallicities, with $\tau$ on the order of 400 Myr.  Since the
best-fit $\tau$ is fairly large, the exponential model is not unlike
the Constant SFR model.  In fact, the two SFH models give similar
masses and ages, which are $\sim 10^{10}$ \Msun\ and $\sim 700$ Myr,
respectively.

{\em Chabrier IMF.} --- In addition to the Salpeter IMF, the
\citet{chabr03} IMF is also implemented in the BC03 models.  The
Chabrier IMF has a flatter distribution than Salpeter at $<$ 1
\Msun. Stars more massive than 1 \Msun\ therefore have a larger
relative contribution to the total luminosity, resulting in a lower
$M/L$ ratio.  Hence, the Chabrier IMF in general produces best-fit
stellar masses that are $\sim 30\% - 40\%$ lower than the Salpeter
IMF.

{\em Alternative Population Synthesis Model.} --- An alternative
population synthesis model was recently introduced by \citet{maras05}.
The main new ingredient of this model is an increased emphasis on the
contributions of thermally pulsating asymptotic giant branch (TP-AGB)
stars to the overall luminosity.  Contributions from TP-AGB stars are
strongest in stellar populations around $\sim 1$ Gyr old, and their
effect is to raise the luminosity at near-IR and longer wavelengths.
In general, this results in a lower $M/L$ ratio, and hence smaller
best-fit masses and ages \citep{mobas05, maras06}.  However, since the
LAEs in our sample are in general much younger than 1 Gyr, and we have
data available only in the rest-frame optical and shorter wavelengths,
we do not expect the \citet{maras05} model to give significantly
different results than the BC03 model.  On the other hand, TP-AGB
stars could potentially contribute to the 5.8 and 8.0 \micron\ flux of
$ z \sim 6$ galaxies, and we stress that it is important to
investigate this alternative model in cases where 5.8 and 8.0 \micron\
data are available.

\section{Comparison to the \imag-dropout Sample} \label{CompLBG}

One of the main goals of this study is to compare and contrast the
properties of the $z \sim 6$ IRAC-detected LAEs and \imag-dropouts.
As we have mentioned in \Sec{Intro}, the different selection criteria
for LAEs and \imag-dropouts may imply fundamental physical differences
between these two populations.  One thing we try to accomplish is to
compare the two populations using model-independent empirical
indicators of the galaxies' properties, such as the rest-frame
UV-to-optical color.  Our sample of IRAC-detected $z \sim 6$
\imag-dropouts is drawn from the works of \citet{dow05},
\citet{eyles05}, and \citet{yan05}.  Other objects are not included
either because they lie at higher redshifts ($z \ga 6.5$) so accurate
$\imag - \zmag$ color cannot be measured owing to Ly$\alpha$ forest
attenuation, or because IRAC photometry is not available.

\subsection{$\imag - \zmag$ color} \label{colorsec}

\begin{figure}
\plotone{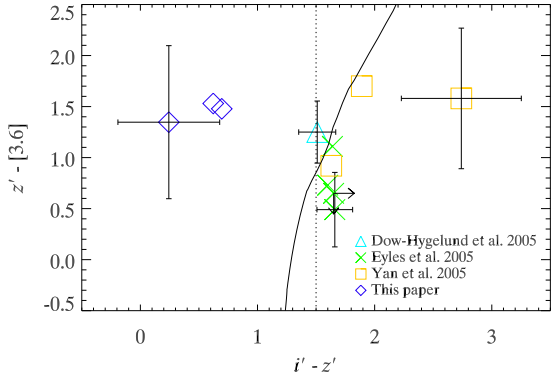}
\caption{$\zmag - [3.6]$ vs.\ $\imag - \zmag$ colors of the LAEs in
  this paper and other IRAC-detected $z \sim 6$ \imag-dropouts in the
  literature.  Only the error bars of the object with the faintest
  \zmag\ magnitude in each group are shown.  This is done to avoid
  clutter and to provide a conservative estimate of the errors
  associated with each group of measurements.  The vertical dotted
  line corresponds to $\imag - \zmag = 1.5$, a typical selection
  criterion for the $z \sim 6$ \imag-dropouts.  The curve shows the
  color evolution of a dust-free, 0.005 \Zsun, SSP mode.  The curve
  corresponds to model ages from $\sim 30$ Myr near the bottom of the
  plot to $\sim 800$ Myr near the top.}
\label{colorplot}
\end{figure}

\begin{figure}
\plotone{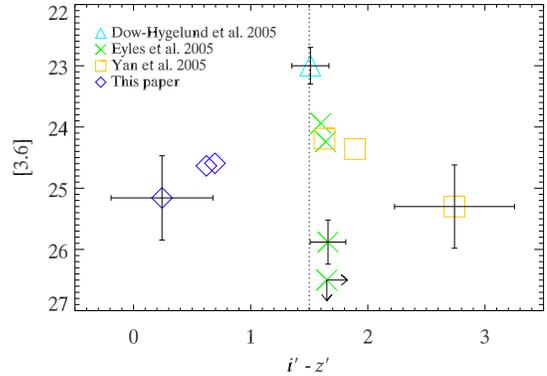}
\caption{IRAC 3.6 \micron\ magnitude vs.\ $\imag - \zmag$ color of the
  LAEs in this paper and other IRAC-detected $z \sim 6$ \imag-dropouts
  in the literature.}
\label{magcolor}
\end{figure}

The $\zmag - [3.6]$ color is plotted against the $\imag - \zmag$ color
in \Fig{colorplot} for both our LAEs and the $z \sim 6$
\imag-dropouts.  The most striking feature in \Fig{colorplot} is that
the LAEs have much bluer $\imag - \zmag$ color than the
\imag-dropouts.  In fact, all three LAEs in our sample have $\imag -
\zmag$ colors that are well blueward of the typical \imag-dropout
selection criterion of $\imag - \zmag > 1.5$.  One possible reason for
this difference is that the \imag-dropouts are preferentially selected
to have red $\imag - \zmag$ color, which may also result in the
\imag-dropouts being selected at slightly higher redshifts than the
LAEs.  Another reason is the unusually high \imag-band flux observed
in our sample (c.f.\ \Sec{Fit}), which may be explained by the
Ly$\alpha$ line contribution.  We have remarked in \Sec{Model} that
the Ly$\alpha$ line can contribute $\sim 30\%$ or more of the total
broad-band flux.  In addition, \citet{hu04} found that, in a sample of
LAEs selected in a similar manner as our objects, there is a clear
correlation between the narrow-band and $I$-band magnitudes, and that
all of the spectroscopically identified LAEs in the sample have
$I-\zmag < 1.5$ (see their Fig.~5).  Similarly, we find that all three
of the LAEs have $\imag - \zmag < 1.5$.

The blue $\imag - \zmag$ colors of the LAEs in our sample implies that
they would have been overlooked by previous surveys based on the
\imag-dropout selection criterion.  It is therefore important to
understand the overlap between the LAE and \imag-dropout galaxies, so
that the properties of the overall $z \sim 6$ galaxy population, such
as the total stellar mass density, can be constrained.

\subsection{Stellar Population} \label{comp_sp}
At $z \sim 6$, the \zmag\ and IRAC 3.6 \micron\ bands sample the
rest-frame UV and optical light, respectively.  If we ignore the
effects of dust extinction, the \zmag\ magnitude is sensitive to the
massive young stars in the galaxy, while the 3.6 \micron\ magnitude
measures the contributions from less massive, evolved stars.  The
$\zmag - [3.6]$ color, plotted against the $\imag - \zmag$ color in
\Fig{colorplot}, is therefore an indicator of the age of the stellar
population.  The model color evolution, also plotted in
\Fig{colorplot}, help to illustrate this point.  The curve starts at
$\sim 30$ Myr near the bottom of the plot.  As the age increases,
$\zmag - [3.6]$ also increases, and eventually reaching $\zmag - [3.6]
= 2.5$ at an age of $\sim 800$ Myr.

As we have mentioned before, the presence of Ly$\alpha$ emission may
signal a young stellar population.  However, \Fig{colorplot} shows
that there is no significant difference between the $\zmag - [3.6]$
color of the LAEs and that of the \imag-dropouts.  Therefore, we find
no evidence to suggest that there is a systematic age difference
between the IRAC-detected LAE and \imag-dropout populations.  However,
we stress again that the $\zmag - [3.6]$ color is a good indicator of
stellar population age only in the absence of dust.

\Fig{magcolor} shows the 3.6 \micron\ band magnitude plotted against
$\imag - \zmag$.  All objects in \Fig{magcolor} lie at similar
redshifts, and hence similar luminosity distances.  This implies the
3.6 \micron\ magnitude may serve as a measure of the total evolved
stellar mass in the galaxies.  The 3.6 \micron\ magnitudes of the LAEs
fall comfortably inside the range of magnitudes observed for the
\imag-dropouts.  Therefore, we again find no evidence that the
IRAC-detected LAEs and \imag-dropouts are systematically different.

In terms of stellar population synthesis modeling, the results are
consistent as well.  For the $z \sim 6$ \imag-dropouts, stellar population
synthesis modeling in general gives ages around a few hundred Myr, and
masses around $10^{10}$ \Msun.  This is mostly in line with the values
we obtained for the LAEs (\Tab{ParTab}), at least in the dust-free SSP
case.

It is important to keep in mind the comparison presented here is
restricted to the IRAC-detected LAEs and \imag-dropouts.  By requiring
IRAC detection for the galaxies, the selection is biased towards the
bright end of the luminosity function.  Recent studies of
\imag-dropouts at $z \sim 6$ suggest that the IRAC-invisible galaxies
are in general younger and less massive \citep{eyles06, yan06}.
\citet{gawis06} also found that LAEs at $z = 3.1$ typically have lower
masses at $\sim \pow{5}{8}$ \Msun.  The similarities seen in
\Fig{colorplot} and \Fig{magcolor} may therefore be a result of this
IRAC-selection bias.

\section{Conclusion} \label{Con}
In this paper, we studied in detail the properties of three Ly$\alpha$
emitting galaxies, each spectroscopically confirmed to lie at $z \sim
5.65$.  Using ACS and IRAC data from GOODS, we measured the galaxies'
SEDs in the rest-frame UV through optical.  Stellar population
synthesis modeling then allows us to place constraints on the
galaxies' masses and ages.  Our main conclusions may be summarized as
follows.

The three IRAC-detected LAEs in our sample exhibit similar emission
properties, characterized by a red rest-frame UV-to-optical color,
with the flux increasing by more than a factor of 3 from \zmag\ to 3.6
\micron.  This feature can be explained by a young stellar population
with significant dust extinction, or an older stellar population that
has developed a Balmer break.  Assuming the SSP model for the star
formation history, we find best-fit stellar populations with masses
between $\sim 10^9 - 10^{10}$ \Msun\ and ages between $\sim 5 - 100$
Myr.  However, stellar populations as old as 700 Myr are admissible if
a constant SFR model is considered.  Very deep near-IR observations
may help to narrow the range of allowed models by providing extra
constraints on the rest-frame UV spectral slope.  The available data
provide very little constraints on the LAEs' metallicity.  We fitted
stellar population synthesis models using five metallicities ranging
from 0.005 \Zsun\ to \Zsun, and found that equally good fits with
similar parameter values can be obtained for all metallicities.

In comparison with other IRAC-detected $z \sim 6$ galaxies selected
based on the \imag-dropout technique, we find that the LAEs and
\imag-dropouts possess similar $\zmag - [3.6]$ colors, suggesting that
they are similar in ages.  Also, the LAEs and \imag-dropouts have
comparable 3.6 \micron\ magnitudes, which imply they have similar
masses.  On the other hand, the comparison is restricted to the
IRAC-detected LAEs and \imag-dropouts, which are likely to be the
brightest and most massive members of their respective populations.
The observed similarities between the IRAC-detected LAEs and
\imag-dropouts may be a result of this selection bias.

Even though the IRAC-detected LAEs and \imag-dropouts share some
common characteristics, the LAEs have much bluer $\imag - \zmag$
colors.  Many previous searches for $z \sim 6$ galaxies were based on
a combination of $\imag - \zmag$ color selection and non-detections in
bands blueward of $\imag$.  Because of their blue $\imag - \zmag$
colors, LAEs would be overlooked by these searches unless additional
selection criteria are incorporated.  One present challenge is
therefore to understand the overlap between the LAE and \imag-dropout
populations.  The solution to this problem will shed light on the $z
\sim 6$ galaxy population in general, and help to constrain the total
stellar mass density, as well as the total contribution of massive
galaxies to the ionization background at $z \sim 6$.

\phantom{}

We would like to thank the referee, Matt Malkan, for insightful
comments that improved the paper.  We also thank Eric Gawiser and
Haojing Yan for helpful discussions.  EMH acknowledges suport from NSF
grant AST06-87850 and LLC from AST04-07374.  This work is based in
part on observations made with the Spitzer Space Telescope, which is
operated by the Jet Propulsion Laboratory, California Institute of
Technology under a contract with NASA. Support for this work was
provided by NASA.

\bibliographystyle{apj}
\bibliography{ms}

\end{document}